\documentclass[aps,prl,twoside,twocolumn,superscriptaddress,floatfix,nofootinbib,showpacs]{revtex4-1}

\usepackage{xcolor}

\usepackage{graphicx}
\usepackage{amssymb, amsmath}
\usepackage[amssymb]{SIunits}
\usepackage{natbib}

\newcommand{\LyA}{\mbox{Lyman-$\alpha$ }}

\newcommand{\sinc}{\mathrm{sinc}}

\newcommand{\Plya}{P_{\mathrm{Ly}\alpha}^{\rm 1d}}

\begin{document}

\title{First Detection of Cosmic Microwave Background Lensing and \LyA Forest Bispectrum}

\author{Cyrille Doux}\email{cdoux@apc.in2p3.fr}
\affiliation{AstroParticule et Cosmologie, Universit\'e Paris Diderot, CNRS, CEA, Observatoire de Paris, Sorbonne Paris Cit\'e \\
B\^atiment Condorcet, 10, rue Alice Domon et L\'eonie Duquet, F-75205 Paris Cedex 13, France}
\author{Emmanuel Schaan}
\affiliation{Dept.~of Astrophysical Sciences, Peyton Hall, Princeton University, Princeton, NJ USA 08544}
\author{Eric Aubourg}
\affiliation{AstroParticule et Cosmologie, Universit\'e Paris Diderot, CNRS, CEA, Observatoire de Paris, Sorbonne Paris Cit\'e \\
B\^atiment Condorcet, 10, rue Alice Domon et L\'eonie Duquet, F-75205 Paris Cedex 13, France}
\author{Ken Ganga}
\affiliation{AstroParticule et Cosmologie, Universit\'e Paris Diderot, CNRS, CEA, Observatoire de Paris, Sorbonne Paris Cit\'e \\
B\^atiment Condorcet, 10, rue Alice Domon et L\'eonie Duquet, F-75205 Paris Cedex 13, France}
\author{Khee-Gan Lee}
\affiliation{Lawrence Berkeley National Laboratory, 1 Cyclotron Road, Berkeley CA 94720, USA}
\affiliation{Hubble Fellow}
\author{David N. Spergel}
\affiliation{Dept.~of Astrophysical Sciences, Peyton Hall, Princeton University, Princeton, NJ USA 08544}
\author{Julien Tr\'eguer}
\affiliation{AstroParticule et Cosmologie, Universit\'e Paris Diderot, CNRS, CEA, Observatoire de Paris, Sorbonne Paris Cit\'e \\
B\^atiment Condorcet, 10, rue Alice Domon et L\'eonie Duquet, F-75205 Paris Cedex 13, France}
%

\begin{abstract}

We present the first detection of a correlation between the \LyA forest and cosmic microwave background (CMB) gravitational lensing.
For each \LyA forest in SDSS-III/BOSS DR12, we correlate the one-dimensional power spectrum with the CMB lensing convergence on the same line of sight from Planck.
This measurement constitutes a position-dependent power spectrum, or a squeezed bispectrum, and quantifies the non-linear response of the \LyA forest power spectrum to a large-scale overdensity. 
The signal is measured at 5~$\sigma$ and is consistent with the $\Lambda$CDM expectation. We measure the linear bias of the \LyA forest with respect to the dark matter distribution, and contrain a combination of non-linear terms including the non-linear bias.
This new observable provides a consistency check for the \LyA forest as a large-scale structure probe and tests our understanding of the relation between intergalactic gas and dark matter.
In the future, it could be used to test hydrodynamical simulations and calibrate the relation between the \LyA forest and dark matter.

\end{abstract}
\pagebreak
\pacs{98.80.-k, 98.70.Vc, 98.62.Ra}
\maketitle

\section{Introduction}

Quasars are bright lanterns that illuminate the dark Universe and probe the distribution of gas. The \LyA forest observed in their spectra reveals the presence of intervening neutral hydrogen absorbing light at $1216\ \angstrom$.
It has been used to study the thermal history of intergalactic gas and hydrogen reionization \cite{2007ApJ...662...72B,2015MNRAS.447.3402B,Lee:2015cm,2011MNRAS.415.2257M}.
Assuming that the flux transmission in the \LyA forest traces the matter density
makes it a powerful probe of the large-scale structure of the Universe at intermediate redshifts, on a wide range of scales.
The one-dimensional power spectrum along the line of sight probes the matter fluctuations on the smallest scales, constraining
the sum of neutrino masses \cite{2015JCAP...11..011P,2015JCAP...02..045P},
models of warm dark matter \cite{2013PhRvD..88d3502V,Seljak:2006hv}
and primordial black hole dark matter \cite{Afshordi:2003ic}.
Combining different lines of sight enables probing the three-dimensional power spectrum on larger scales \cite{2011MNRAS.415.2257M,2011JCAP...09..001S}
and provides a measurement of the baryonic acoustic oscillations (BAO) at high redshift \cite{Slosar:2013bz,2013A&A...552A..96B,2015A&A...574A..59D}.

The interpretation of these 
results rely on the \LyA transmission tracing the underlying matter density field.
If the hydrogen is in photoionization equilibrium with a uniform UV background, and there is no other sources of entropy, then the relationship
is described analytically through variations of the fluctuating Gunn-Peterson approximation (FGPA; \cite{Croft:1998gi}) and is evaluated numerically using hydrodynamical simulations \cite{2014JCAP...07..005B,Rossi:2014jy}.
However, the connection between \LyA transmission and the underlying matter density is complex \cite{2014PhRvD..89h3010P} and non-linear. It is affected by the proximity effect on the largest scales \cite{2014PhRvD..89h3010P}, by thermal broadening, Jeans smoothing and non-linear gravitational evolution on the smallest scales \cite{2015JCAP...12..017A} and by the gas equation of state throughout.
For these reasons, and in the light of the tension between BAO measurements from the \LyA forest and galaxies \cite{2015PhRvD..92l3516A}, consistency checks for the link between \LyA transmission and matter density are valuable.

Gravitational lensing of the cosmic microwave background (CMB) is sourced by large-scale structures located between the last scattering surface and the observer, and provides a measurement of the projected density of matter.
In this paper, we cross-correlate, for the first time,
the power spectrum of 
the flux transmission in the \LyA forest of quasars with CMB lensing to test our understanding of the relation between intergalactic gas and dark matter.

\begin{figure}[t]\centering\includegraphics[width=1\columnwidth]{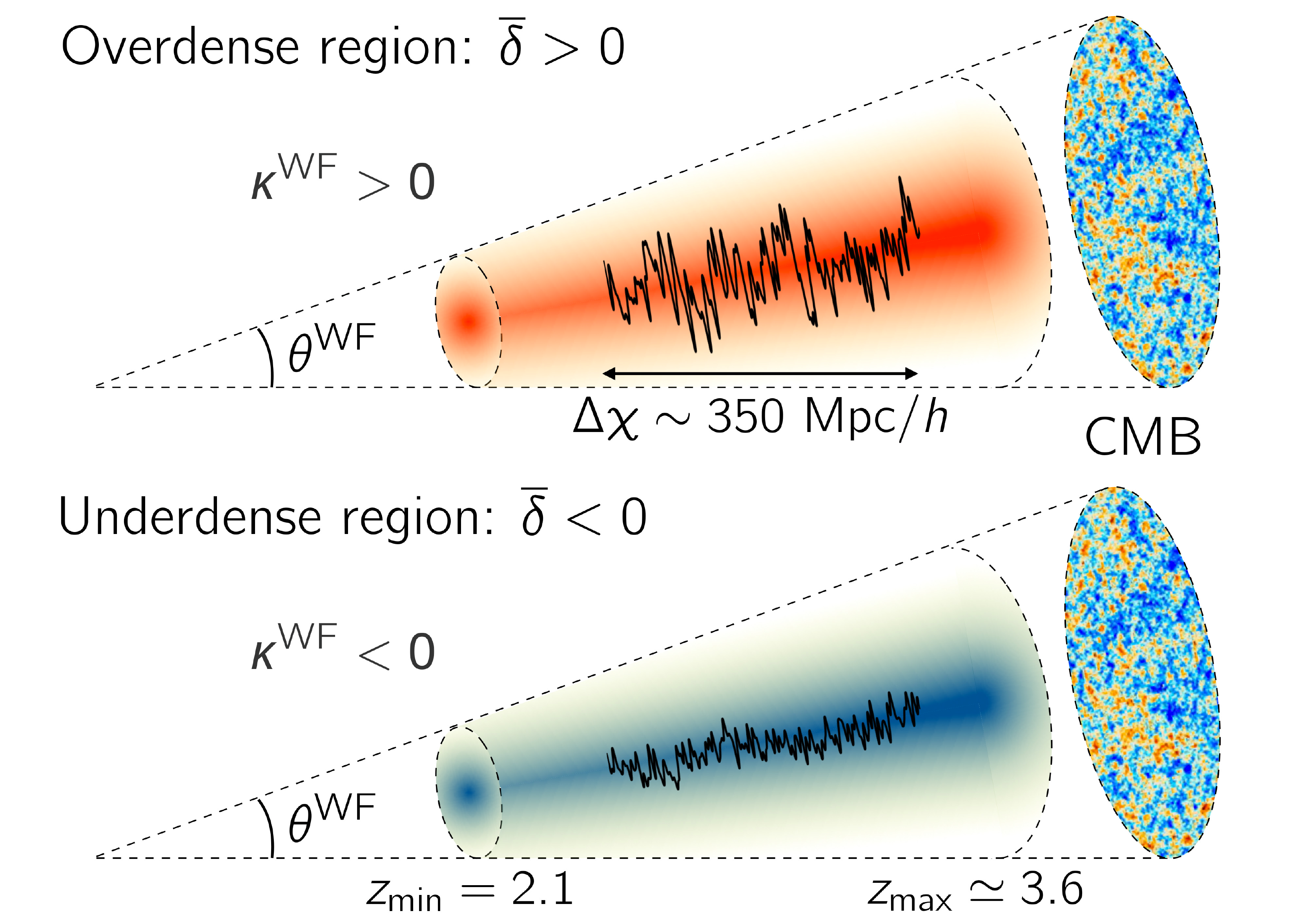}\caption{Schematic of the correlation: overdense regions (respectively underdense regions), in red on the top panel (blue on the bottom panel) have positive (negative) CMB lensing convergence and are expected to produce more (less) small-scale structures under non-linear gravitational evolution, which is detectable in the amplitude of the \LyA forest power spectrum. The extent of the aforementioned regions is determined by the angular resolution $\theta^{\rm WF}$ of the Wiener-filtered convergence map and depth of the lensing efficiency function. In this analysis, we select \LyA forests in the redshift range $2.1-3.6$.}\label{fig:schematic}\end{figure}

Since both the CMB lensing convergence and the mean \LyA transmission probe the mean density on a given line of sight, it is natural to compute their cross-spectrum. However, the mean \LyA transmission is strongly affected by continuum fitting in the quasar spectrum, making this a challenging observable.
For this reason, we instead correlate the CMB lensing convergence with the small-scale \LyA power spectrum on the same line of sight. The origin of this signal is more complex, and corresponds to a position-dependent power spectrum \cite{2014JCAP...05..048C,2014PhRvD..90l3523S}, or a squeezed bispectrum of the matter density.
Simply put, a positive CMB convergence corresponds to an overdense line of sight; on this line of sight, the matter power spectrum is enhanced on all scales, due to non-linear evolution under gravity \cite{2014JCAP...05..048C,Li:2014hm,2014PhRvD..90l3523S} (see Fig.~\ref{fig:schematic} for a schematic of this idea). This bispectrum would therefore vanish at linear order in the perturbation theory of the density field, where short and long modes are independent. Instead, for a non-linear density field, this signal probes the response of the \LyA power spectrum to a mean overdensity. 

This method was proposed in \cite{2001ApJ...551...48Z,2009PhRvL.103i1304V,Vallinotto:2011bn}. In this paper, we present the first detection of this signal, and propose a new theoretical description of it, based on the response of the matter power spectrum to a mean overdensity.

\section{Theory}

We aim at evaluating the covariance between the one-dimensional power spectrum $\Plya(k_\parallel)$ of the \LyA forest transmission on one line of sight and the CMB convergence $\kappa$ on the same line of sight.

The CMB lensing convergence $\kappa$ probes the large-scale matter distribution along the line of sight. Our estimate $\hat{\kappa}$ comes from CMB lensing reconstruction \cite{2002ApJ...574..566H,Okamoto:2003bu} and is Wiener-filtered such that $\hat{\kappa} = \Lambda_{\kappa} \ast \kappa + \text{noise}$.
As a result, the estimated convergence effectively probes the mass distribution within a ``cone'', whose line of sight dimension is determined by the lensing efficiency kernel ${W_\kappa(\chi) = \delta \kappa / \delta \left[\delta(\chi)\right]}$, and whose angular size is determined by the Wiener filter $\Lambda_{\kappa}$. This is depicted in Fig.~\ref{fig:schematic}.

We split this cone into thin slices of fixed comoving distance $\chi$ and thickness $d\chi$.
The variance of the density field $\bar{\delta}(\chi)$ averaged over this thin slice is given by $\mathrm{Var}\left[ \bar{\delta}(\chi) \right]= \sigma^2(\chi) / d\chi$ (see \cite{2014PhRvD..90l3523S}) with
\begin{equation}\sigma^2(\chi)=\int \frac{d^2 k_\perp}{\left( 2\pi \right)^2}\left| \Lambda_{\kappa}(\ell = \chi k_\perp) \right|^2P_\text{lin}(k_\perp, \chi)\end{equation}
where $P_\text{lin}(k_\perp, \chi)$ is the linear matter power spectrum at comoving distance $\chi$.
The covariance of $\bar{\delta}(\chi)$ with the 1d power spectrum measured on the same slice is \cite{2014PhRvD..90l3523S}
\begin{equation}\text{Cov}\left[\bar{\delta}(\chi),\Plya(k_\parallel,\chi)\right]=\text{Var}\left[\bar{\delta}(\chi)\right]\;\frac{\partial\Plya(k_\parallel,\chi)}{\partial\delta}+\mathcal{O}(\sigma^4).\label{eq:response_p1d_delta}\end{equation}
In other words, the response of the \LyA power spectrum to the mean matter overdensity produces a non-zero covariance. This is the signal we wish to detect.
For a given \LyA forest, measured between $\chi_\mathrm{min}$ and $\chi_\mathrm{max}$, we define an average power spectrum
\begin{equation}\Plya(k_\parallel)=\frac{1}{\Delta\chi}\int_{\chi_\mathrm{min}}^{\chi_\mathrm{max}}d\chi\Plya(k_\parallel,\chi)\end{equation}
where ${\Delta\chi = \chi_\mathrm{max} - \chi_\mathrm{min}}$.
Since the CMB convergence $\kappa = \int d\chi W(\chi) \delta(\chi)$ is a weighted average of the matter density field, 
we perform the same average on Eq.~\eqref{eq:response_p1d_delta} to get an integrated bispectrum between the CMB lensing convergence and fluctuations in the \LyA forest
\begin{eqnarray}B_{\kappa,{\mathrm{Ly}\alpha}}(k_\parallel)&\hat{=}&\text{Cov}\left[\kappa,\Plya (k_\parallel)\right]\nonumber\\
&=&\frac{1}{\Delta\chi}{\int}d\chi{W}_\kappa(\chi)\frac{\partial\Plya(k_\parallel,\chi)}{\partial\delta}\sigma^2(\chi),\label{eq:response}\end{eqnarray}
where the integral runs from $\chi_\mathrm{min}$ to $\chi_\mathrm{max}$. We assumed the various redshift slices are uncorrelated, as in the Limber approximation \cite{1953ApJ...117..134L,2008PhRvD..78l3506L}, because the lensing window function $W_\kappa(\chi)$ varies slowly over the integral width, which in turn, is much larger than the scales where the correlation is expected, of order $0.1-1\ h / \mathrm{Mpc}$.

To go further, we need to evaluate the response of the \LyA power spectrum to the mean overdensity $\partial \Plya / \partial \delta$. The 1d power spectrum is related to the 3d power spectrum as
\begin{equation}\Plya(k_\parallel)=\int\frac{d^2\vec{k}_\perp}{(2\pi)^2}\;P^{3d}_{\mathrm{Ly}\alpha}(k_\parallel,\vec{k}_\perp),\label{eq:p1d_th}\end{equation}
which, in turn, is related to the linear matter power spectrum as \cite{2015JCAP...12..017A}
\begin{equation}P^{3d}_{\mathrm{Ly}\alpha}(k_\parallel,\vec{k}_\perp)=b_1^2\left(1+\beta\mu^2\right)^2D(k,\mu)P_\text{lin}(k),\label{eq:p3d_th}\end{equation}
where $\mu$ is the cosine of the angle between $\vec{k}_\parallel$ and $\vec{k}=\vec{k}_\parallel+\vec{k}_\perp$.
The term $b_1$ represents the linear bias of the \LyA transmission, $\beta$ corresponds to linear redshift-space distortions, and $D$ encapsulates several non-linearities: Jeans smoothing at small scales under gas pressure, thermal broadening of absorption lines due to local thermal velocity dispersion and finally non-linear structure formation under gravity (it goes to 1 as $k$ goes to zero and has an exponential cut-off at small scales).
This fitting formula was obtained in \cite{2015JCAP...12..017A} by comparing hydrodynamical simulations of the intergalactic medium and assumes that the ionizing UV background is homogeneous. It neglects possible fluctuations in this background or in the density-temperature relation arising from inhomogeneous reionization, as well as astrophysical effects such as galactic winds and quasar effects which would most likely affecting only very small volumes (see \cite{2015JCAP...12..017A} and references therein).

In the presence of an overdensity $\delta$, the linear power spectrum responds as \cite{2014JCAP...05..048C}
\begin{equation}\frac{\partial{\ln}P_\text{lin}}{\partial\delta}=\frac{68}{21}-\frac{1}{3}\frac{\partial{\ln}k^3P_\text{lin}}{\partial{\ln}k}.\end{equation}
The linear bias term $b_1$, the Kaiser term $\beta$ and the baryonic non-linearities encapsulated in $D(k,\mu)$ may also respond to a mean overdensity. We characterize the response of these terms by an effective non-linear bias
\begin{equation}b_2^{\mathrm{eff}}(k,\mu)=\frac{\partial}{\partial\delta}\ln\left(b_1^2\left(1+\beta\mu^2\right)^2D(k,\mu)\right).\label{eq:b2eff}\end{equation}
This quantity combines non-linear bias\footnote{Note that the first-order non-linear bias $b_2$ is generally defined through the response of the linear bias to an overdensity via $\partial b_1/\partial\delta = b_2 - b_1^2$, \emph{e.g.} in \cite{2014JCAP...05..048C,2016JCAP...09..007B}. Therefore, if we neglect the Kaiser term  and the baryonic term, the quantity $b_2^{\mathrm{eff}}$ defined here is related to the non-linear bias $b_2$ as $b_2^{\mathrm{eff}}=2(b_2-b_1^2)/b_1$.} and the response of redshift-space distorsions and non-linear clustering of gas. It will be measured from the bispectrum, and can, in principle, be measured from simulations.
The response of the \LyA forest power spectrum is thus
\begin{equation}\frac{\partial\Plya(k_\parallel)}{\partial\delta}=\int\frac{d^2\vec{k}_\perp}{(2\pi)^2}\;P^{3d}_{\mathrm{Ly}\alpha}(\vec{k})\left(\frac{\partial{\ln}P_\text{lin}}{\partial\delta}+b_2^{\mathrm{eff}}(k,\mu)\right).\label{eq:resp_p1d}\end{equation}
Combining Eqs~\eqref{eq:response} and \eqref{eq:resp_p1d}, the CMB lensing - \LyA bispectrum becomes the sum of two terms:
one representing the response of the linear matter power spectrum, and one for the non-linear and baryonic terms.

\section{\LyA forest power spectrum}

We use quasar spectra from the twelfth data release of {SDSS-III/BOSS}~\cite{2012AJ....143...51L, 2013AJ....145...10D,Paris:2016wa}. The continuum fitting is performed using a mean-flux-regulated principal component analysis method described in \cite{2012AJ....143...51L} that was applied to the DR9 in \cite{2013AJ....145...69L}. The domain of the \LyA forest of a given quasar spectrum is defined by limits on the rest frame wavelength
\begin{equation}1041\angstrom\leq\lambda_{\mathrm{rf}}\leq1185\angstrom,\end{equation}
where $\lambda_{\rm rf} = \lambda / (1+z_{\rm QSO})$ for a quasar at redshift $z_{\rm QSO}$.

Spectra displaying damped \LyA absorption systems (DLAs, identified using the technique described in \cite{Noterdaeme:2012fm}) or broad absorption lines (BALs, identified by visual inspection in \cite{Paris:2016wa}) were discarded.
We select quasars with a signal-to-noise ratio in the \LyA forest, measured by the BOSS pipeline, greater than 1 and a redshift between 2.15 and 4.0. The noise estimation gives poor results close to the spectrograph blue-end, so we additionally cut parts of the forests below $z=2.1$ as in \cite{2013A&A...559A..85P}. Finally, we discard quasars lying outside of the Planck lensing mask. These cuts select 87,085 quasars out of the 155,002 in the DR12 catalog.

The flux transmission fraction in the \LyA forest at redshift $z=\lambda / \lambda_{\mathrm{Ly}\alpha}$ is
\begin{equation}F(z)=\frac{f(z)}{C(z)},\end{equation}
where $f(z)$ is the measured flux and $C(z)$ is the estimated continuum. We then estimate the normalized transmitted flux fraction as a function of redshift as
\begin{equation}\delta_i(z)=\frac{F_i(z)}{{\langle}F(z){\rangle}}-1\end{equation}
where $\langle F(z) \rangle$ is the mean flux transmission fraction obtained by averaging over quasars and $i$ stands for the forest index.

The normalized flux fraction is converted from a function of redshift to a function of radial comoving distance $\chi(z)$, which is evaluated using cosmological parameters from Planck 2015 (TT,TE,EE+lowP+lensing+ext) \cite{Ade:2015xua}. Because the spacing between pixels of the BOSS spectrograph is logarithmic in wavelength, with ${\Delta(\log_{10} \lambda) = 10^{-4}}$, and because sky emission lines are masked (on average 1.2\% of pixels), the spacing in distance space is slightly irregular, albeit monotonically growing.
Therefore the Fourier transform $\tilde{\delta} (k)$ of the normalized transmitted flux fraction $\delta(\chi)$ is computed using the NFFT library \cite{Keiner:2009js}.
For a forest of length $\Delta \chi$ and mean pixel spacing $\overline{\delta\chi}$, the smallest mode is $k_{\rm min} = 2\pi / \Delta \chi$ while the largest mode is $k_{\rm max} = 2\pi / \overline{\delta\chi}$. Forests have a mean length of {$420-280$ Mpc/$h$} with a spacing of order {$0.68-0.58$ Mpc/$h$} for forests in the redshift range ${2.1-3.6}$, giving ${k_{\rm max} \approx 6\ h/{\rm Mpc}}$. These scales are highly affected by non-linear clustering and baryonic effects. Moreover, the power spectrum becomes noisier because of the resolution of the spectrograph (see the spectrograph window function in Eq.~\eqref{eq:wspectro}), so we restrict our analysis to ${k_{\rm max} = 1.5\ h/{\rm Mpc}}$, corresponding to scales of order {4 Mpc/$h$}, consistent with \cite{2013A&A...559A..85P}.
We note that the large-scale modes over ${60\ \mathrm{Mpc}/h}$, \emph{i.e.} ${k \lesssim 0.1\ h/{\rm Mpc}}$ may be slightly affected by continuum fitting, though in a way uncorrelated with CMB lensing.
The raw power spectrum is obtained by
\begin{equation}\hat{P}_{i}^{\mathrm{raw}}(k)=\frac{|\tilde{\delta}_i(k)|^2}{\Delta\chi_i}\end{equation}
where $\Delta \chi_i$ is the comoving length of the $i$-th. forest.

Multiple observations of the same quasars allow for an estimation of the noise level $\sigma_{\rm noise}^2(\chi)$ for each pixel in the forest. We assume the noise to have a white spectrum, which agrees well with \cite{2013A&A...559A..85P,2011MNRAS.415.2257M}, and estimate its power spectrum by averaging
\begin{equation}P_i^{\mathrm{noise}}=\overline{\sigma_{\mathrm{noise},i}^2}\frac{\pi}{{\Delta}k}\end{equation}
where $\Delta k = k_{\rm max} - k_{\rm min}$.

The resolution of the spectrograph is of order {1 Mpc/$h$} and varies slowly with wavelength (by about 10\% over one forest). Therefore the spectrograph window function is
\begin{equation}W_{\mathrm{spectro}}(k,R_i)=\exp\left(-\frac{k^2R^2_i}{2}\right)\times\sinc\left(\frac{k\overline{\delta\chi}_i}{2}\right)\label{eq:wspectro},\end{equation}
where $R_i$ is the resolution of the spectrograph averaged over the $i$-th forest,
\begin{equation}R_i=\overline{\frac{c(1+z)}{H(z)}\delta_{\mathrm{disp}}\Delta\log\lambda},\end{equation}
with $\delta_{\rm disp}$ being the measured dispersion in units of $\Delta \log_{10} \lambda$. The second term, in which $\sinc(x) = \sin(x)/x$, accounts for the pixelization.
Finally, the estimator of the one-dimensional power spectrum of the \LyA forest is given by
\begin{equation}\hat{P}_{\mathrm{Ly}\alpha}^{\mathrm{1d}}(k,z)=\left\langle\frac{\hat{P}_{i}^{\mathrm{raw}}(k)-P_i^{\mathrm{noise}}}{W_{\mathrm{spectro}}^2(k,R_i)}\right\rangle_{i{\in}z}.\label{eq:P1D_est}\end{equation}
where the average is over forests falling in the redshift range.

This straightforward measurement of the one-dimensional power spectrum does not correct for subtle instrumental systematics dealt with in \cite{2013A&A...559A..85P}. However, these effects are less important in a cross-correlation measurement, and a precise estimation of the power spectrum is beyond the scope of this paper.

\section{CMB lensing}

We use the publicly available\footnote{See \url{http://pla.esac.esa.int/pla/}.} CMB lensing convergence map from Planck 2015 \cite{Ade:1987168}. We apply a Wiener filter to the map using the provided signal and noise power spectra to get the minimum-variance linear estimator for the CMB convergence $\kappa^{\rm WF}\left( \mathbf{\theta}_i\right)$ in the direction of each \LyA forest.

\section{Results \& interpretation}
\begin{figure}[t]\centering\includegraphics[width=1\columnwidth]{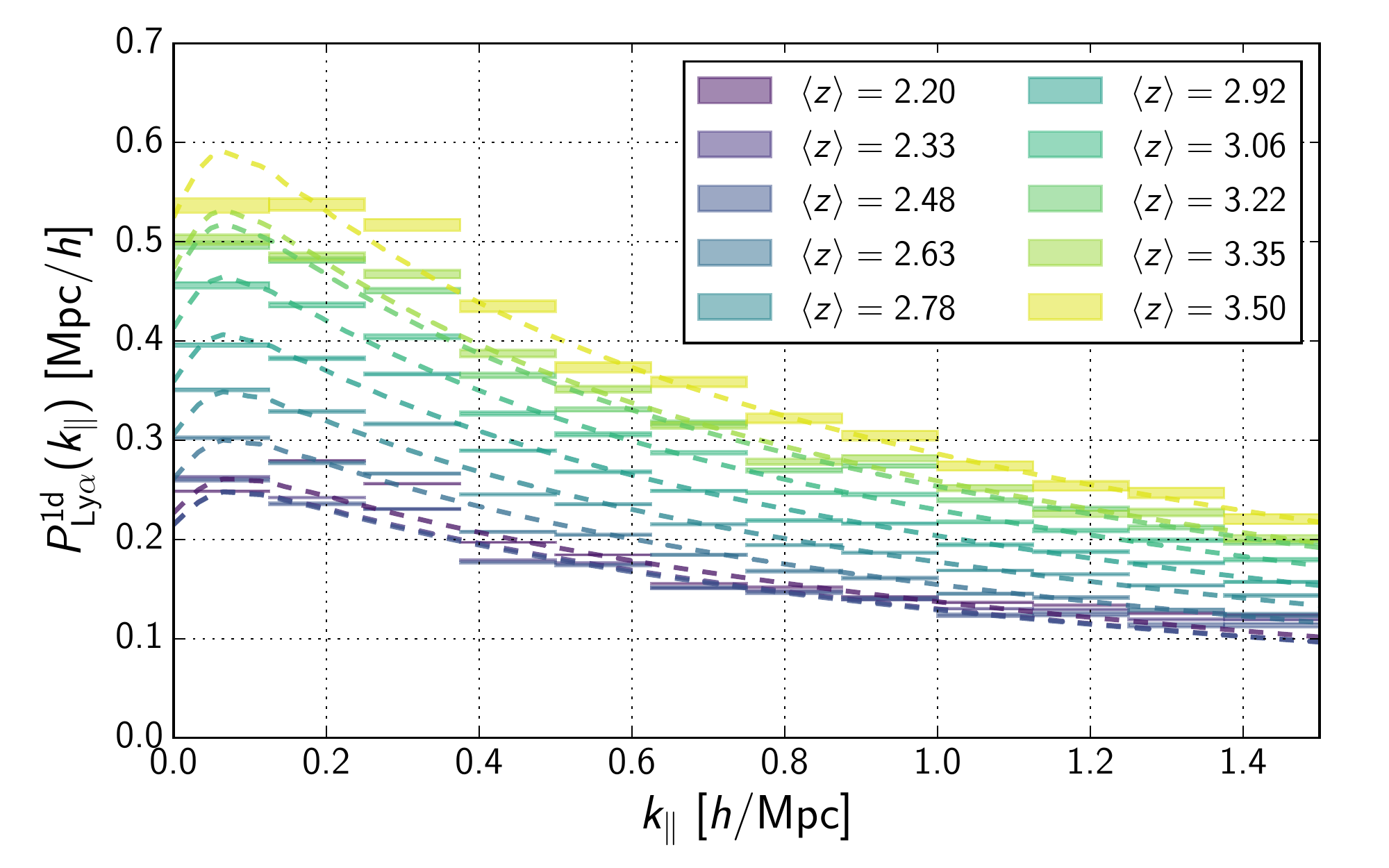}\caption{One-dimensional power spectrum of the \LyA forest $P_{\mathrm{Ly}\alpha}^{\mathrm{1d}}$ in 10 redshift bins. The redshifts indicate the mean value of the middle redshifts of the forests. Error bars are computed from the weighted empirical covariance of the power spectra of different forests. Colored boxes represent the measured spectra averaged over these redshifts bins with their uncertainties and $k$-bins width. Dashed lines are theoretical curves with fitted bias $b_1(z)$.}\label{fig:p1d}\end{figure}
\begin{figure}[t]\centering\includegraphics[width=1\columnwidth]{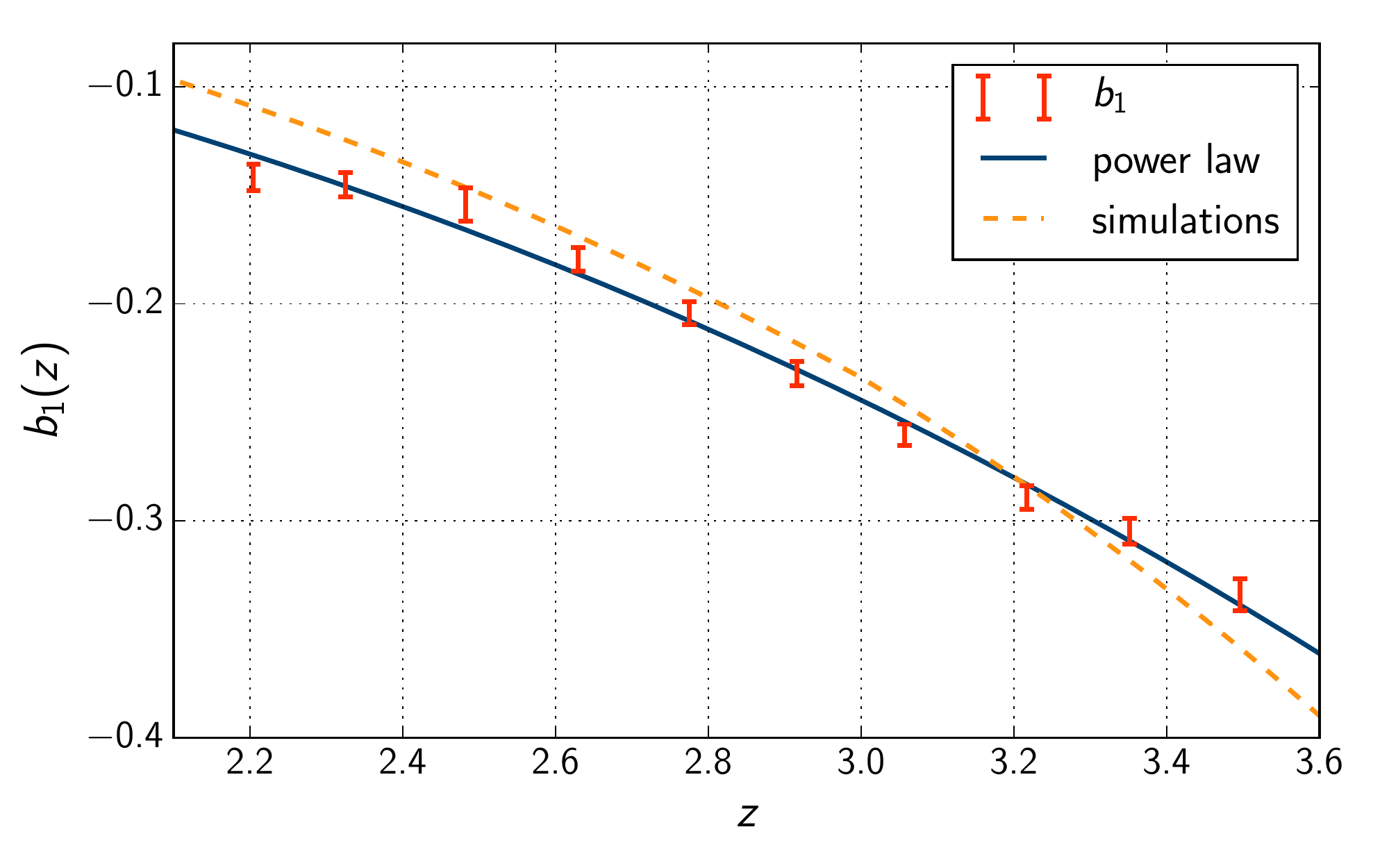}\caption{\LyA flux transmission linear bias $b_1$ as a function of redshift. The data points are depicted with error bars and the power law in ${(1+z)}$ (see text) is represented by the blue solid curve. The dashed orange curve represents the linear bias measured from hydrodynamical simulations in \cite{2015JCAP...12..017A}. Remark that the transmission is theoretically anti-correlated with the amount of hydrogen, hence the negative sign.}\label{fig:bias}\end{figure}
The first step of our analysis is to measure the one-dimensional power spectrum in order to obtain the linear bias $b_1(z)$ from Eq.~\eqref{eq:p3d_th} as a function of redshift. The theoretical curves are computed using Eqs.~\eqref{eq:p1d_th} and \eqref{eq:p3d_th} with parameters, except for the linear bias that we aim at fitting, measured from simulations in \cite{2015JCAP...12..017A}. We divide our forest sample into 10 linearly-spaced redshift bins using the central redshift of each forest. Each power spectrum is given a scale-dependent minimum-variance weight
\begin{equation}w_i(k){\propto}n_i^{\mathrm{pix}}\left(P_{\mathrm{Ly}\alpha}^{\mathrm{1d,fid}}(k,z_i)+\frac{P_i^{\mathrm{noise}}}{W_{\mathrm{spectro}}^2(k,R_i)}\right)^{-2},\label{eq:weights}\end{equation}
where $n_i^{\rm pix}$ is the number of pixels of the $i$-th forest. The fiducial power spectrum $P_{\mathrm{Ly}\alpha}^{\rm 1d,fid}(k,z)$ is computed using the linear bias measured from simulations in \cite{2015JCAP...12..017A}, which will only be used in the weights.

In order to take into account possible wavelength-dependent bias in the noise estimation, we allow for a common rescaling of the estimated noise power spectrum in each redshift bin. Precisely, we introduce a coefficient $\alpha_z$ in front of $P_i^{\rm noise}$ in Eq.~\eqref{eq:P1D_est}, common to all spectra in each redshift bin and fit this parameter jointly with the linear bias $b_1(z)$ in each redshift bin. The estimated one-dimensional power spectrum is shown in Fig.~\ref{fig:p1d} together with theoretical curves with best fit biases. The best fit of the linear bias $b_1(z)$ is shown in Fig.~\ref{fig:bias} with error bars including the marginalization over $\alpha_z$. We fit this result with a power law in ${(1+z)}$ of the form
$b_1(z) = a(1+z)^b$
and find $a=-0.00507$ and $b=2.79$. It is represented by the solid blue curve in Fig.~\ref{fig:bias} and is in fairly good agreement with the bias measured in hydrodynamic simulations in \cite{2015JCAP...12..017A} (which we only used in the weights).

\begin{figure}[t]\centering\includegraphics[width=1\columnwidth]{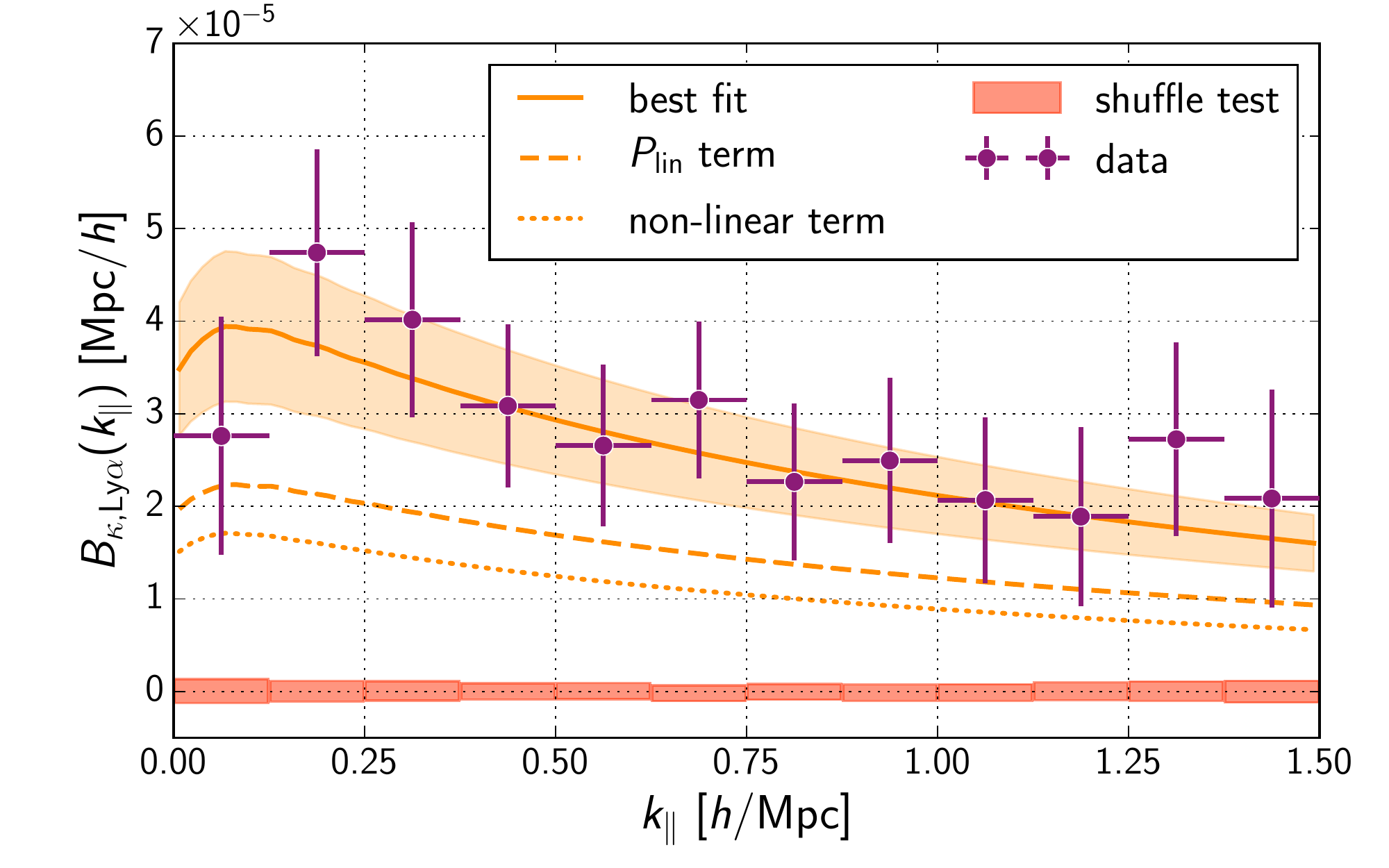}\caption{Integrated bispectrum of CMB lensing and fluctuations in the \LyA forest. The Wiener-filtered CMB lensing is measured in the direction of quasars for which we measure the \LyA forest one-dimensional power spectrum in the range $k_{\parallel}\sim0.1-1.5\ h/\mathrm{Mpc}$. Data points (in purple) show a signal measured at 5~$\sigma$. The theoretical curve (solid orange) is the sum of two terms: the response of the linear matter power spectrum (dashed), and the response of the non-linear terms in the \LyA power spectrum (non-linear bias $b_2$, Kaiser term and baryonic non-linear term $D$) (dotted). While the first involves no free parameter, the latter has an amplitude characterized by the effective non-linear bias $b_2^{\rm eff} = 1.16 \pm 0.53$, see Eq.~\eqref{eq:b2eff}. The orange area represents the 1~$\sigma$ uncertainty on this non-linearity amplitude. We test that our estimator is coherent with zero in the case of no correlation by a shuffling method (thin red boxes, expanded 10 times for visibility).}\label{fig:response}\end{figure}

The next step of our analysis is to compute the weighted unbiased covariance of the lensing convergence and the one-dimensional power spectrum. Quasars have a significant contribution to the lensing of the CMB because the lensing efficiency $W_\kappa$ peaks at $z\sim2$. Therefore, we expect the mean convergence in the directions of quasars to be positive, and indeed find ${10^4 \times \langle {\kappa_i^{\rm WF}} \rangle = 1.35 \pm 0.52}$. This value is consistent with the expected amplitude $\kappa = (\Lambda_{\kappa} \ast \Sigma ) / \bar{\rho} \sim 1.5 \times 10^{-4}$ where $\Sigma$ is the projected density of the haloes hosting the quasars (computed for a NFW profile \cite{Navarro:1996ce} with a halo mass $M_{\rm h} \sim 2 \times 10^{12} M_{\odot} / h$ and redshift 2.5 \cite{2012MNRAS.424..933W}) convolved with the Wiener filter and $\bar{\rho}$ is the mean matter density.
With the aim of measuring the correlation between our two probes, we subtract the mean value ${\langle {\kappa_i^{\rm WF}} \rangle}$ in the computation of the covariance. So as to decrease the effects of noise in this measurement, we also subtract the mean value of the power spectrum in each $k$-bin.
The estimator for the correlation of CMB lensing and fluctuations in the \LyA forest, \emph{i.e.} the ${\mathrm{CMB \ lensing}-\mathrm{Ly}\alpha}$ integrated bispectrum, is defined as
\begin{equation}\hat{B}_{\kappa,{\mathrm{Ly}\alpha}}(k_\parallel)\hat{=}\mathrm{Cov}_{w(k_\parallel)}\left[\kappa^{\mathrm{WF}},\Plya(k_\parallel)\right],\label{eq:signal}\end{equation}
where
\begin{equation}\mathrm{Cov}_{w}\left[x,y\right]=\mathcal{N}\times\sum_{i}w_i\left(x_i-{\langle}x{\rangle}\right)\left(y_i-{\langle}y{\rangle}\right)\end{equation}
with the normalization ${\mathcal{N} = \sum_i w_i / \left( \left(\sum_i w_i\right)^2 - \sum_i w_i^2 \right)}$.
The mean values $\langle {\kappa^{\rm WF}} \rangle$ and $\langle {P^{\rm 1d}(k_\parallel)} \rangle$ are computed using the same weights as well. The measured values in each $k$-bin are shown in purple in Fig.~\ref{fig:response}.

\begin{figure}[t]\centering\includegraphics[width=1\columnwidth]{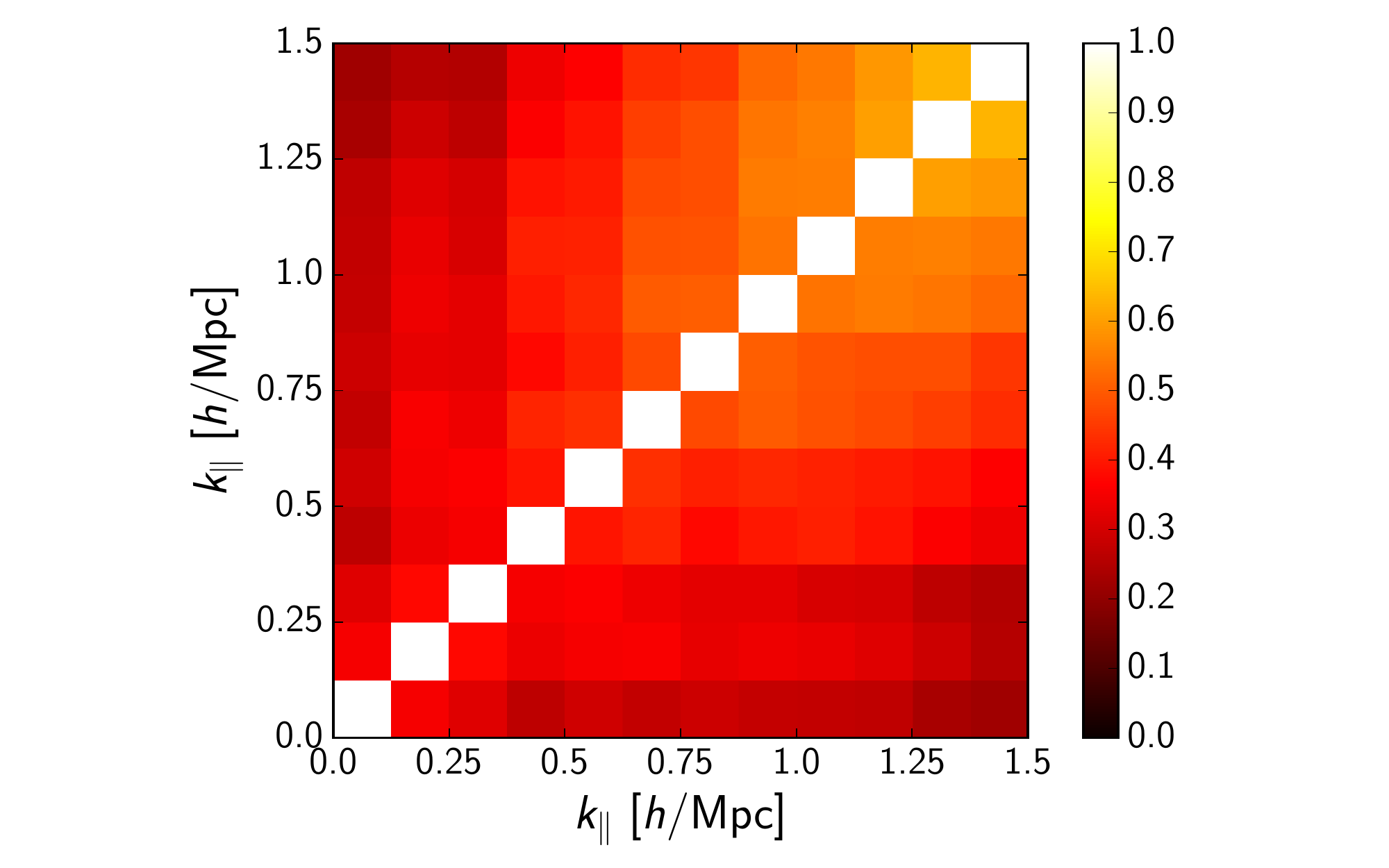}\caption{Correlation matrix of the data-points between $k$-bins computed by shuffling the indices of one of the variables. It shows an important correlation ranging from 20\% up to almost 65\% for the large $k$ modes.}\label{fig:bigcov}\end{figure}

To compute the covariance matrix for the various $k$-bins,
we proceed by computing the signal repeatedly with shuffled indices in $\kappa_i^{\rm WF}$. More precisely, for a given random permutation $\sigma$ of the quasar indices, we compute $\text{Cov}\left[ \kappa_{\sigma(i)}^{\rm WF}, P^{\rm 1d}_{\mathrm{Ly}\alpha,i}(k_\parallel) \right]$ and repeat $N=10,000$ times. We then estimate the mean value (thin red boxes on Fig.~\ref{fig:response}) and the empirical covariance. 
The corresponding matrix of correlation coefficients is shown in Fig.~\ref{fig:bigcov}.

Finally, we aim at comparing our theoretical model and fitting a value of the effective non-linear bias $b_2^{\rm eff}$ defined in Eq.~\eqref{eq:b2eff}. We measure a single number, \emph{i.e.} a scale and redshift averaged non-linear bias integrated over $\mu$, characterizing the non-linear response in our sample. This parametrization is incomplete, but sufficient given the signal-to-noise ratio. For each line of sight, we evaluate the expected signal using Eq.~\eqref{eq:response} given the redshift range $\left[ z_{\rm min}, z_{\rm max}\right]$ of the forest, and the linear bias $b_1(z)$ from the power law best fit.
We then weight the theoretical expected value by the weights in Eq.~\eqref{eq:weights}. The best fit value is $b_2^{\rm eff} = 1.16 \pm 0.53$. The theoretical curve (in orange in Fig.~\ref{fig:response}) is the sum of two contributions, one from the linear power spectrum (dashed line) and the other from the non-linear terms (dotted line).

Using the covariance matrix obtained by our shuffling method and the measured data points, we find a $\chi^2$ value for the null hypothesis $\chi^2_{\rm null} = 30.1$ for 12 data points. The probability to exceed is 0.27\% and the null hypothesis is therefore rejected at a significance of 3.0~$\sigma$. For the best fit in $b_2$, we find $\chi^2_{\rm best-fit} = 5.37$, a small value that could be explained by over-estimated error bars, which would lead to a better detection. The signal-to-noise ratio for the detection is ${\rm SNR} = \sqrt{\chi^2_{\rm null} - \chi^2_{\rm best-fit} } = 4.97$, hence this constitutes a 5~$\sigma$ detection of the non-linear response of the \LyA power spectrum.

\section{Null tests}

In order to assess the cosmological nature of this signal, we proceed to a number of null tests.

First, we make sure that the correlation estimator is consistent with zero in the case of no correlation. To do so, we compute the mean of the values of the signal measured with shuffled indices in $\kappa^{WF}_i$ as we expect different lines to be uncorrelated. The result is consistent with zero, as shown on Fig.~\ref{fig:response} by the thin red boxes.

\begin{figure}[t]\centering\includegraphics[width=1\columnwidth]{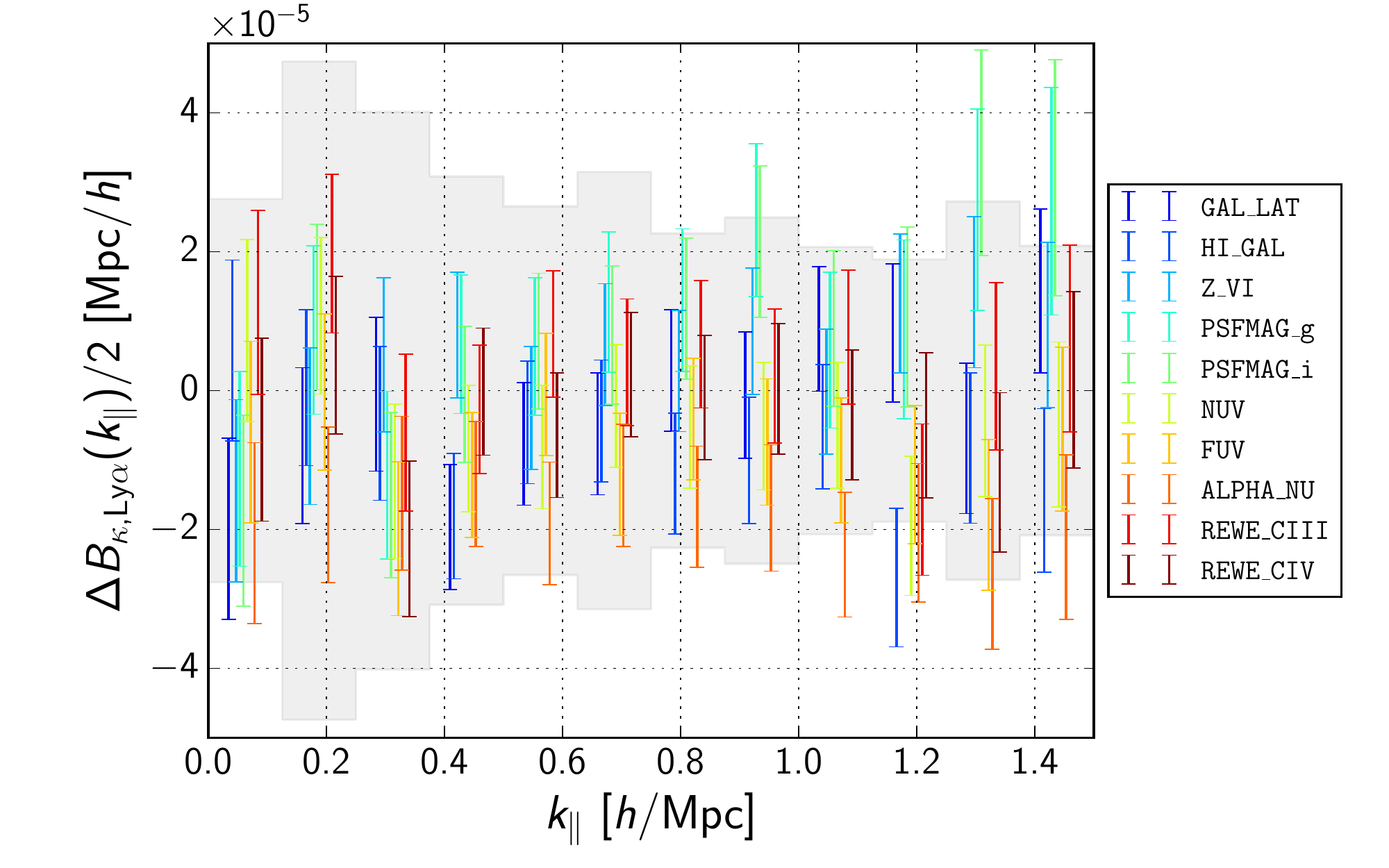}\caption{Null tests for various quasar properties. For each parameter, the sample is split in two parts according to the median value. We then compute the ${\mathrm{CMB \ lensing}-\mathrm{Ly}\alpha}$ integrated bispectrum for these two sub-samples, using Eq.~\eqref{eq:signal} and compute the difference (divided by 2 in order to have the same error bars as the signal itself). The greyed area is delimited by the absolute level of the measured bispectrum for the full sample (as in Fig.~\ref{fig:response}).}\label{fig:null_tests}\end{figure}

Second, we want to verify that the signal we measure does not originate in a possible correlation of the lensing convergence in the directions of quasars with their intrinsic properties. We split our sample in two equal parts according to the median values of various quasar parameters. For each of these parameters, we measure the signal in the two sub-samples using Eq.~\eqref{eq:signal} and compute the difference. The results are shown in Fig.~\ref{fig:null_tests} and the meaning of the tested parameters' names are detailed in Table~\ref{tab:params}. We test for galactic latitude of the quasars and galactic H\textsc{i} column density in their directions to verify that the signal is not related to galactic foregrounds. We test for quasar redshift and find no significant variation of the signal. We also test for various intrinsic properties of the quasars linked to their masses: colors (PSF magnitudes in the $g$ and $i$ bands), near and far UV fluxes (from GALEX \cite{2011MNRAS.413.2570R}) and quasar spectral index. We also test for contamination by carbon lines using the rest equivalent width of the emissions of C\textsc{iii} and C\textsc{iv}. We find that all tests are consistent with zero at the ${0.2-1.6\ \sigma}$ level. Lack of data prevents us from testing the contamination from the Si\textsc{iv} line; however, it is at most a 5\% effect according to \cite{2013A&A...559A..85P}.

\begin{table}
	\caption{\label{tab:params} Results of the null tests seeking for correlation between the lensing signal and intrinsic quasars properties.}
	\begin{ruledtabular}
	\begin{tabular}{lll}
		Label & Description & Null test \\
		\hline
		\texttt{GAL\_LAT}    & Galactic latitude & 1.6~$\sigma$\\
		\vspace{0.25cm}
		\texttt{HI\_GAL}     & log of galactic H\textsc{i} column density & 1.1~$\sigma$\\
		\vspace{0.25cm}
		\texttt{Z\_VI}       & Quasar redshift from visual inspection & 0.2~$\sigma$\\
		\texttt{PSFMAG\_g}   & PSF magnitude (flux in $g$ band) & 1.0~$\sigma$\\
		\texttt{PSFMAG\_i}   & PSF magnitude (flux in $i$ band) & 1.4~$\sigma$\\
		\texttt{NUV}         & Near UV flux (from GALEX) & 0.7~$\sigma$\\
		\texttt{FUV}         & Far UV flux (from GALEX) & 0.5~$\sigma$\\
		\vspace{0.25cm}
		\texttt{ALPHA\_NU}   & Quasar spectral index & 0.5~$\sigma$\\
		\texttt{REWE\_CIII}  & Rest equivalent width of C\textsc{iii} emission & 1.1~$\sigma$\\
		\texttt{REWE\_CIV}   & Rest equivalent width of C\textsc{iv} emission & 0.3~$\sigma$\\
	\end{tabular}
	\end{ruledtabular}
\end{table}

\section{Discussion}

We have presented the first detection of a cross-correlation between the \LyA forest of quasars and the gravitational lensing of the CMB. Our understanding of this correlation is based on the response of small-scale fluctuations in the matter density, measured by the one-dimensional power spectrum of the transmission in the \LyA forest, to large-scale overdensities probed by the convergence of CMB lensing. This signal corresponds to a bispectrum in the squeezed limit configuration where the two small-scale modes are of order $k \sim 0.1 - 1\ \mathrm{Mpc}/h$. This is the first measurement of the ${\mathrm{CMB \ lensing}-\mathrm{Ly}\alpha}$ integrated bispectrum, and it measures the non-linearity in the \LyA forest. Finally, this new observable tests our understanding of the relation between neutral hydrogen and dark matter.

We measured the one-dimensional power spectrum and the linear bias of the \LyA forest, finding values consistent with hydrodynamical simulations. Even though the power spectrum is sensitive to a number of systematic effects, these are much less important in a cross-correlation measurement like the integrated bispectrum that we computed. The theoretical bispectrum is the sum of two contributions: the response of the linear matter power spectrum, theoretically well-understood and involving no free parameters, and the response of the bias and non-linear terms, computed up to an effective non-linear bias $b_2^{\rm eff}$ which we have fitted. We believe this model provides a reasonable explanation of the observed signal.

However, we notice that our interpretation of the measured bispectrum is limited by theoretical uncertainties mainly related to baryonic physics. That is, the term $D(k, \mu)$, taken from \cite{2015JCAP...12..017A}, encodes a number of effects that are significant at very small scales (of order $k\sim 60\ h/\mathrm{Mpc}$, see \cite{2016JCAP...03..016C}), but the integral of the three-dimensional power spectrum in Eq.~\eqref{eq:p1d_th} gets contributions from $k$-modes greater than 10~$h$/Mpc, and we cannot neglect these effects. Moreover, this term and the redshift-space distorsion term $\beta$ may also respond to large-scale overdensities. Therefore, the effective non-linear bias term $b_2^{\rm eff}$ encompasses several uncertain contributions: it could be compared with simulations, providing both a valuable check for the simulation assumptions while shedding light on the relation between \LyA and dark matter.

Another uncertainty arises from the fact that the ionizing UV flux of the quasars reduces the amount of neutral hydrogen around them, a phenomenon known as the proximity effect. Because overdense regions radiate more, the bias of neutral hydrogen $b_{\mathrm{H\textsc{i}}}(k)$ becomes negative at scales larger than ${k\sim0.01\ h/\mathrm{Mpc}}$ \cite{2014PhRvD..89h3010P}. This impacts the H\textsc{i} power spectrum for scales ${k \lesssim 0.1\ h/\mathrm{Mpc}}$ and may also affect our measurement in the lowest $k$-bin.

In the future, hydrodynamical simulations could be compared to this new observable and used to model the dependence of the various non-linear terms in the one-dimensional power spectrum on the mean overdensity. In particular, the anisotropic linear bias of the \LyA forest, \emph{i.e.} its dependence on angle $\mu$, could be explored. Meanwhile, this measurement can inform simulations and help contrain the relation between intergalactic gas and dark matter. Other avenues of exploration are to study the dependence of the signal on redshift and perpendicular separation $r_\perp$. The upcoming {SDSS-IV/eBOSS} \cite{2016AJ....151...44D} data, covering a broad redshift range, will improve the signal-to-noise ratio and help measure the redshift dependence. Combined with a high signal-to-noise ratio CMB lensing map, it could allow for a measurement of the angular correlation between the \LyA forest and large-scale overdensities.
Finally, because this bispectrum is sensitive to small scales observed in the \LyA forest, it could also provide additional constraints on the total mass of neutrinos and be used as a tool to study alternative models of dark matter predicting small-scale cut-offs.

\section*{Acknowledgments}

	The authors warmly thank the referee for interesting and useful questions and comments.
	
	C.D., E.S. and E.A. acknowledge support from CNRS grant PICS APC-Princeton. C.D., E.S. and D.N.S. acknowledge support from NSF grant no. AST-1311756. K.-G.L. acknowledges support for this work by NASA through Hubble Fellowship grant HF2-51361 awarded by the Space Telescope Science Institute, which is operated by the Association of Universities for Research in Astronomy, Inc., for NASA, under contract NAS5-26555..
	
	Funding for SDSS-III has been provided by the Alfred P. Sloan Foundation, the Participating Institutions, the National Science Foundation, and the U.S. Department of Energy Office of Science. The SDSS-III web site is http://www.sdss3.org/.

	SDSS-III is managed by the Astrophysical Research Consortium for the Participating Institutions of the SDSS-III Collaboration including the University of Arizona, the Brazilian Participation Group, Brookhaven National Laboratory, Carnegie Mellon University, University of Florida, the French Participation Group, the German Participation Group, Harvard University, the Instituto de Astrofisica de Canarias, the Michigan State/Notre Dame/JINA Participation Group, Johns Hopkins University, Lawrence Berkeley National Laboratory, Max Planck Institute for Astrophysics, Max Planck Institute for Extraterrestrial Physics, New Mexico State University, New York University, Ohio State University, Pennsylvania State University, University of Portsmouth, Princeton University, the Spanish Participation Group, University of Tokyo, University of Utah, Vanderbilt University, University of Virginia, University of Washington, and Yale University.

\bibliographystyle{prsty}
\bibliography{paper_lya_cmblens}

\end{document}